\documentstyle[12pt,epsf]{article}

\begin{document}
\title{$4\pi$ Decay Modes of the $f_0(1500)$ Resonance} 
\author{M. Strohmeier-Pre\v{s}i\v{c}ek, T. Gutsche, Amand Faessler\\
{\small\it Institut f\"ur Theoretische Physik, Universit\"at T\"ubingen},\\
{\small\it Auf der Morgenstelle 14, 72076 T\"ubingen, Germany}\\
and\\
R. Vinh Mau\\
{\small\it Division de Physique Th\'{e}orique, Institut de Physique Nucl\'{e}aire, 
91406 Orsay Cedex, France} \\
{\small\it and LPTPE, Universit\'{e} P. et M. Curie, 4 Place Jussieu, 75230 Paris Cedex,
France}}
\date{}
\maketitle
\begin{abstract}
We investigate the two-body decay modes $\rho\rho ,~\pi\pi^{\ast}(1300)$ and
$\sigma \sigma$ of the $f_0(1500)$, all leading to the $4\pi$ decay channel,
in a three-state mixing scheme, where the $f_0(1500)$ is a mixture 
of the lowest-lying scalar glueball with the nearby isoscalar states of the 
$0^{++}\, Q\bar{Q}$ nonet. In the leading order of this scheme,
the decay mechanism of the $f_0(1500)$ proceeds
dominantly via its quarkonia components, which can be described 
in the framework of the $^3P_0$ $Q\bar Q$ pair
creation model.
We predict the hierarchy of decay branching ratios B with 
$B(\rho \rho ) ^{>}_{\sim} B(\pi\pi ) ^{>}_{\sim} B(\sigma \sigma )
> B( \pi\pi^{\ast} )$,
providing a key signature for the proposed mixing scheme
in this leading order approach.

\end{abstract}
\bigskip
\bigskip


\newpage

Recently, there has been growing attention, both experimental and theoretical,
on the possible  existence of glueball states.
A scalar-isoscalar resonance $f_0(1500)$ is clearly
established by Crystal Barrel at LEAR \cite{crys,anis94} in
proton-antiproton annihilation, which
is regarded as a rich gluon production
mechanism with favored coupling to glueballs \cite{clos}.
The main interest in the $f_0(1500)$ as a candidate reflecting the
intrusion of a glueball state in the scalar-isoscalar meson spectrum
rests on
several phenomenological and theoretical observations.
Lattice QCD, in the quenched approximation \cite{bal},
predicts the lightest glueball to be a scalar, lying
in the mass range of 1500-1700 MeV.
Although the decay pattern of the $f_0(1500)$ into two pseudoscalar
mesons is compatible with
a quarkonium state $n\bar{n} \equiv \frac{1}{\sqrt{2}}(u\bar{u}+d\bar{d})$
in the scalar $Q\bar Q$ nonet, the observed total width of $\Gamma
\approx 120~MeV$  is in conflict 
with naive quark model expectations of $\Gamma \ge 500~MeV $
\cite{kok}.
Moreover, it is now believed \cite{crys,ams},
that the $f_0(1370)$ state occupies the corresponding position
in the scalar nonet, with decay branching ratios and total width consistent
with a dominant $n\bar n$ structure.
As the $f_0(1500)$ cannot be accommodated as a radial
excitation of the scalar $^3P_0$ nonet \cite{barn}, 
it appears as an excess state in the $Q\bar{Q}$
systematics.

In the analysis of the partial decay widths into
two pseudoscalar mesons, several schemes have been proposed to
attribute at least
a partial glueball nature to the $f_0(1500)$ state \cite{ams,ani,wein},
although quantitative predictions for its glueball content differ
sizably.
In the scenario proposed by  Ref. \cite{ams} the pure glueball
is mixed with  the  $n\bar n$ and $s\bar s$ states  of
the scalar $^3P_0$ nonet.
The mixture leads to three scalar states with nearly equal glueball strength,
identified with the
$f_0(1370)$, $f_0(1500)$ and an additional state around 1700 MeV
which can possibly be  the $f_{J=0,2}(1710)$.
It is argued that in the leading order of strong coupling 
QCD, the hadronic two meson decay proceeds dominantly via the $Q\bar{Q}$-
components of the $f_0(1500)$, if the Fock states of final state isoscalar
mesons contain no gluonic component.
Neglecting an intrinsic violation
of flavor symmetry, the model can explain the decay pattern
of the $f_0(1500)$ into two pseudoscalar mesons.
In particular, the experimental weak $K\bar K$ decay mode 
is obtained  from the destructive interference between  the admixed
$n\bar n$ and $s\bar s$ modes.

There are  hints that the scalar $s\bar s$ state
may have a lower mass than the scalar glueball motivating another
three-state mixing model with this level ordering
\cite{wein}.
In this scenario the glueball component dominantly resides in the
$f_J(1710)$ with the $f_0(1500)$ remaining as a nearly pure quarkonium state,
but the destructive interference
between the $n\bar n$ and $s\bar s$
flavors for the $f_0(1500)$ still remains.
For this reason, both analyses of Refs. \cite{ams} and \cite{wein} are
very similar for the decay of $f_0(1500)$
into two  pseudoscalar mesons \cite{ams}.

In the present work, we go beyond the two 
pseudoscalar meson decay 
modes and  study the predictions of  the three-state
 mixing schemes  of Refs.
\cite{ams,wein} in  the $4\pi$ decay modes.
We derive the decay pattern of the $f_0(1500)$ into the two-body
decay channels  $\rho\rho ,~\pi\pi^{\ast}(1300)$ and $\sigma \sigma$
($\sigma$ is the broad  $\pi\pi$ S-wave resonance).
They  all  contribute  to  the $4\pi$
decay rate, and we calculate their relative 
 ratios to that of the  $2\pi$ decay channel.
The two-body decay modes leading to the $4\pi$ final state have the
advantage that,  in the decay dynamics,  no $SU(3)$ flavor symmetry breaking
and  mixing angles for the meson nonets in the decay channel
are needed.
This reduces the uncertainties occuring  in the analysis of the two
pseudoscalar meson decays.

{\it The model} - As mentioned earlier,  the hadronic decay
of the  admixed $f_0(1500)$ is domi\-nated by the $Q\bar Q$ components.
The coupling of the scalar 
$Q\bar{Q}$ components to the final two-meson states can be calculated 
in the non-relativistic $Q\bar{Q}$ pair creation $^3P_0$ model
\cite{le},
in which the  $Q\bar{Q}$ pairs are created  with vacuum quantum numbers.
The relevant OZI-allowed quark line diagram, describing the 
flavor flux from the initial $Q\bar Q$ component of the $f_0(1500)$
to the two meson final states BC (=$\pi\pi ,~\rho\rho ,~\sigma\sigma$ and
$\pi\pi^{\ast}(1300)$) is shown in the figure.
The details of all the following formulae can be found in Refs.
\cite{kok,barn,le,ackley}.
The transition dynamics is governed by the nonperturbative $Q\bar Q$
$^3P_0$ vertex:
\begin{equation}
V_{^3P_0}^{(34)}= \lambda \delta ( {\mathbf{p}} _{4}+ 
{\mathbf{p}} _{3}) \left[ {\cal Y}_{1\mu}^\ast( {\mathbf{p}} _{4}
- {\mathbf{p}} _{3}) \otimes \sigma_{-\mu}^{(43)\dagger} \right]_{00}  
{\mathbf{1}} _F^{(43)} {\mathbf{1}} _C^{(43)}~,
\end{equation}
 ${\mathbf{p}} _{3(4)}$ are the momenta of the created 
$Q\bar Q$ pair and ${\cal Y}_{1\mu}^\ast ({\mathbf{p}})=|{\mathbf{p}}|
Y_{1\mu} ^\ast (\hat{ {\mathbf{p}}})$.
The identity operators ${\mathbf{1}}_F$ and ${\mathbf{1}}_C$ project onto
singlet states in flavor and color space, respectively. 
The dimensionless parameter $\lambda$ corresponds to the
strength of the transition, which in turn is related to the probability for
$Q\bar Q$ pair creation out of the hadronic vacuum.  
Since for the considered final states BC
no $s\bar s$ pair creation is involved
$\lambda$ is flavor independent.
With harmonic oscillator wave functions for the quark
clusters, the scalar $Q\bar Q$-component of the $f_0(1500)$ is given as:
\begin{eqnarray} \label{a}
\Psi_{f_0}({\mathbf{p}}_1,{\mathbf{p}}_2) & = & \psi_{S_{f_0}=1 ,
l_{f_0}=1 , J_{f_0}=0}
({\mathbf{p}}_1,{\mathbf{p}}_2) \chi_F^{f_0}\chi_C^{f_0} \nonumber\\
 & = & (-i) \left[ \frac{2R_{f_0}^5}{3\sqrt{\pi}} \right]^{\frac{1}{2}} \delta ( {\mathbf{p}}_1 + {\mathbf{p}}_2 )
\exp \left\{ - \frac{1}{8} R_{f_0}^2 ( {\mathbf{p}}_1-{\mathbf{p}}_2)^2 \right\}\\ 
 &  & \cdot \left[ \chi_{S=1}(12) \otimes {\cal Y}_{l=1}({\mathbf{p}}_1-{\mathbf{p}}_2) \right]_{J=0} \chi_{F}^{f_0} \chi_C^{f_0} ~, \nonumber
\end{eqnarray}
 $R_{f_0}$ is a size parameter,
the intrinsic spin $\chi_{S=1}$ of the $Q\bar Q$ pair
is coupled with relative orbital angular momentum $l=1$ to 
the $J^{PC}=0^{++}$ state,
 $\chi_C$  denotes the color wave function and
the effective flavor part $\chi_F^{f_0}$ is given by 
$\chi_F^{f_0}={\xi}\, (n\bar{n})$, 
where $\xi$ represents the amplitude for the $n\bar n$ component in
the three-state mixing scheme.
In the same way, for the final state mesons $\sigma$ and $\pi^\ast$
(which is the first radial excitation of the $\pi$ with $n_{\pi^\ast}=1$)
with decay momentum ${\mathbf{K}}$ and intrinsic quark momenta
${\mathbf{p}} _{i(j)}$, the wave functions are
\newpage
\begin{eqnarray} \label{b}
\Psi_{\sigma}({\mathbf{p}}_i,{\mathbf{p}}_j) & = & \psi_{S_{\sigma}=1 ,
l_{\sigma}=1 ,J_{\sigma}=0}({\mathbf{p}}_i ,{\mathbf{p}}_j) \chi_{F}^{\sigma} \chi_{C}^{\sigma}
 \nonumber\\
& = & (-i) \left[ \frac{2 R_\sigma^5}{3\sqrt{\pi}} \right]^{\frac{1}{2}} 
\delta ( {\mathbf{p}}_i+{\mathbf{p}}_j - {\mathbf{K}}) \exp \left\{ -\frac{1}{8}
R_{\sigma}^2 ( {\mathbf{p}}_i-{\mathbf{p}}_j)^2 \right\} \\
 &  & \cdot \left[ \chi_{S_{\sigma}=1}(ij) \otimes {\cal Y}_{l_{\sigma}=1}
(\mathbf{p}_i-\mathbf{p}_j) \right]_{J_{\sigma}=0}
\chi_{F}^{\sigma} \chi_{C}^{\sigma} ~,  \nonumber
\end{eqnarray}
\begin{eqnarray}\label{c}
\Psi_{\pi^\ast}({\mathbf{p}}_i ,{\mathbf{p}}_j) & = &
 \psi_{n_{\pi^\ast}=1,S_{\pi^\ast}=0,
l_{\pi^\ast}=0,J_{\pi^\ast}=0}({\mathbf{p}}_i , {\mathbf{p}}_j)
 \chi_{F}^{\pi^\ast}\chi_{C}^{\pi^\ast}  \nonumber\\
 & = & (-) \left[\frac{8R_{\pi^\ast}^3}{3\sqrt{\pi}} \right]^{\frac{1}{2}} 
\delta({\mathbf{p}}_i+{\mathbf{p}}_j-{\mathbf{K}})\exp \left\{ -\frac{1}{8}R_{\pi^\ast}^2 ({\mathbf{p}}_i-{\mathbf{p}}_j)^2 \right\} \\
 &  & \cdot L_{n_{\pi^\ast}=1}^{1/2}(\frac{R_{\pi^\ast}^2}{4}({\mathbf{p}}_i-
{\mathbf{p}}_j)^2) \left[ \chi_{S_{\pi^\ast}=0}(ij) \otimes Y_{l_{\pi^\ast}=0}
(\hat{{\mathbf{p}}_i-{\mathbf{p}}_j}) \right]_{J_{\pi^\ast}=0} \chi_{F}^{\pi^\ast}\chi_{C}^{\pi^\ast}
  \nonumber
\end{eqnarray}
with the Laguerre polynomial $L_{n=1}^{1/2} (p) = 3/2 -p$.
Here, $\sigma $ is effectively described by an isoscalar $^3P_0$ $Q \bar Q$
state.  
The S-wave mesons $\pi$ and $\rho$ are described by
\begin{eqnarray} \label{d}
\Psi_{\pi(\rho)}({\mathbf{p}}_i,{\mathbf{p}}_j) & = & 
\psi_{S_{\pi(\rho)},l_{\pi(\rho)}=0,J_{\pi(\rho)}}({\mathbf{p}}_i,
{\mathbf{p}}_j)
\chi_{F}^{\pi(\rho)}\chi_{C}^{\pi(\rho)}   \nonumber\\
 & = & \left[ \frac{4 R_{\pi(\rho)}^3}{\sqrt{\pi}} \right]^{\frac{1}{2}}
 \delta ( {\mathbf{p}}_i+{\mathbf{p}}_j-{\mathbf{K}}) 
\exp \left\{ - \frac{1}{8} R_{\pi(\rho)}^2 ({\mathbf{p}}_i-{\mathbf{p}}_j)^2 \right\} \\
 &  &  \cdot \left[ \chi_{S_{\pi(\rho)}}(ij) \otimes Y_{l_{\pi(\rho)}=0}
(\hat{{\mathbf{p}}_i-{\mathbf{p}}_j}) \right]_{J_{\pi(\rho)}} 
\chi_{F}^{\pi(\rho)}\chi_{C}^{\pi(\rho)}  \nonumber
\end{eqnarray} 
with $S_\pi=J_\pi=0, S_\rho=J_\rho=1$.
The size parameters for $\pi$ and $\pi^{\ast}$ have to fulfill the
relation $R_{\pi}=R_{\pi^\ast}$ due to orthogonality of the wave functions.

The transition amplitude for the  process 
$f_0(1500) \rightarrow BC$ of the  figure  is given by
\begin{eqnarray}
T & = & < \Psi_B^{(13)}({\mathbf{p}}_{1},{\mathbf{p}}_{3})
\Psi_C^{(42)}({\mathbf{p}}_{4}, {\mathbf{p}}_{2})|
V_{^3P_0}^{(34)}
|\Psi_{f_0}({\mathbf{p}}_1, {\mathbf{p}}_2)>  \\ 
& = & T_{SS}^{f_0\rightarrow BC}
T_F^{f_0 \rightarrow BC} T_C^{f_0 \rightarrow BC} \nonumber
\end{eqnarray}
The   color part  $T_C$ is  $ \frac{1}{\sqrt{3}}$,  and the
 spin-spatial part $T_{SS}^{f_0 \rightarrow BC}$ is given with the
appropriate wave functions of Eqs. (2)-(5) as 
\begin{equation}
T_{SS}^{f_O \rightarrow BC}  =  \int \prod_{i=1,...,4}
d {\mathbf{p}}_i 
 \psi_{S_Bl_BJ_B}^{\dagger}( {\mathbf{p}}_{1},{\mathbf{p}}_{3})
 \psi_{S_Cl_CJ_C}^\dagger ({\mathbf{p}}_{4},{\mathbf{p}}_{2})
 V_{^3P_0}^{(34)}
\psi_{S_{f_0}=1,l_{f_0}=1,J_{f_0}=0}({\mathbf{p}}_1,{\mathbf{p}}_2)~,
\end{equation}
where the internal quark momenta are integrated over and evaluation of the
integral (7) is done analytically \cite{le}.
The various flavor factors $T_F^{(f_0 \rightarrow BC)}$ are indicated in the
table.
Using relativistic phase space,  the partial decay width
$\Gamma_{f_0 \rightarrow BC}$ is written as
\begin{equation} \label{ph}
\Gamma_{f_0 \rightarrow BC} = 2\pi \frac{E_BE_C}{M_{f_0}} \, K
\int d \Omega_K | T({\mathbf{K}})|^2 
\end{equation}
with decay momentum $|{\mathbf{K}}|=K$ and where
$E_i=\sqrt{M^2_i + {\mathbf{K}}^2 }$
and $M_i$ (i=B,C) are energy and mass of the outgoing meson i.
Analytical evaluation of Eq. (\ref{ph}) leads to  
\begin{eqnarray}
\Gamma_{f_0\rightarrow \sigma\sigma} & = & \lambda^2 \frac{E_\sigma^2}{M_{f_0}}
\, \frac{16}{81 \sqrt{\pi}}\, \frac{R_{f_0}^5 R_\sigma^{10}}{(R_{f_0}^2+2R_\sigma^2)^7} \, \exp \left\{ - \frac{1}{2} \left( \frac{R_{f_0}^2 R_{\sigma}^2}{
R_{f_0}^2+2R_{\sigma}^2} \right) K^2 \right\} \nonumber\\
 & & \cdot  | T_F^{f_0 \rightarrow \sigma\sigma} |^2 \,K
\left[ 60  
 +\frac{11R_{f_0}^4+4R_\sigma^2(R_{f_0}^2+R_\sigma^2)}{(R_{f_0}^2+
2R_\sigma^2)} K^{2} \right. \\
& & \left.  - \frac{R_{f_0}^4
R_\sigma^2 (R_{f_0}^2+R_\sigma^2)}{(R_{f_0}^2+2R_\sigma^2)^2} K^{4}
 \right]^2 ~, \nonumber
\end{eqnarray} 
\begin{eqnarray}
\Gamma_{f_0 \rightarrow \pi^\ast\pi} & = & \lambda^2  \frac{E_{\pi^\ast}E_\pi}
{M_{f_0}} \, \frac{8}{27\sqrt{\pi}}\, \frac{R_{f_0}^5 R_{\pi}^6}
{(R_{f_0}^2+2R_{\pi}^2)^7} \exp \left\{ -\frac{1}{2} \left( 
\frac{R_{f_0}^2 R_{\pi}^2}{R_{f_0}^2+2R_{\pi}^2} \right) K^2 \right\} \nonumber\\
& & \cdot | T_F^{f_0 \rightarrow \pi^\ast\pi} |^2 \, K
\left[ (18R_{f_0}^2-24R_{\pi}^2)   
 - \frac{R_{f_0}^2R_{\pi}^2(13R_{f_0}^2+6R_{\pi}^2)}{(R_{f_0}^2+2R_{\pi}^2)}
 K^{2} \right. \nonumber\\
& & \left. + \frac{R_{f_0}^4R_{\pi}^4 (R_{f_0}^2+R_{\pi}^2)}
{(R_{f_0}^2+2R_{\pi}^2)^2} K^{4} \right]
^2  
\end{eqnarray}
and
\begin{eqnarray}
\Gamma_{f_0\rightarrow \pi\pi(\rho\rho)} & = & \lambda^2 \, \frac{E_{\pi(\rho)}
^2}{M_{f_0}}\, \frac{64}{A\sqrt{\pi}}\, \frac{R_{f_0}^5R_{\pi(\rho)}^6}{(R_{f_0}^2+2R_{\pi(\rho)}^2)^5} \exp \left\{ -\frac{1}{2} \left( \frac{R_{f_0}^2
R_{\pi(\rho)}^2}{R_{f_0}^2+2R_{\pi(\rho)}^2} \right) K_{\pi(\rho)}^2 \right\} \nonumber\\
& & \cdot | T_F^{f_0 \rightarrow \pi\pi} |^2\, K_{\pi(\rho)}
 \left[ 1 - \frac{2}{3}\, \left(
\frac{R_{\pi(\rho)}^2 (R_{f_0}^2+R_{\pi(\rho)}^2)}
{R_{f_0}^2+2R_{\pi(\rho)}^2}\right) K_{\pi(\rho)}^2
\right. \\ 
& & \left. + \, \frac{1}{B} \left( \frac{R_{\pi(\rho)}^2(R_{f_0}^2+R_{\pi(\rho)}^2)}
{R_{f_0}^2+2R_{\pi(\rho)}^2} \right)^2 K_{\pi(\rho)}^4  \right] \nonumber
\end{eqnarray}
with $A=1, B=9$ for $f_0 \rightarrow \pi\pi$ and $A=3, B=1$ for 
$f_0 \rightarrow \rho\rho$.
For the broad mesons $f_0(1500), \pi^\ast (1300), \rho(770)$ and $\sigma(760)$
we average over the mass spectrum $f(m)$:
\begin{eqnarray}
\Gamma_{f_0 \rightarrow BC} & = & \int dm_{f_0} dm_B dm_C ~
 \Gamma_{f_0 \rightarrow BC}(m_{f_0}, m_B, m_C)
f(m_{f_0}) f(m_B) f(m_C) \\
f(m) & \propto & \frac{(\Gamma_i/2)^2}{(m-M_i)^2+(\Gamma_i/2)^2} \nonumber
\end{eqnarray}
with  a proper threshold cutoff  introduced as in Ref. \cite{ma}.
Masses $M_i $ and widths $ \Gamma_i$ are taken from the Particle Data Group.
They are well defined except $\Gamma_{\pi^\ast}$ which is 
$(400\pm200)\, MeV$. We parameterize the mass distribution of  
 the scalar-isoscalar $\sigma$ meson as in Ref. \cite{nag}
($M_\sigma =$ 760 MeV, $\Gamma_\sigma=$ 640 MeV).
The partial decay widths of all $4\pi$ channels and also the $2\pi$ mode
scale with $(\xi \cdot \lambda)^2$. 
The  predictions for relative branching ratios are therefore 
 independent of
any  particular three-state mixing scheme \cite{ams,wein},
specified by  the explicit value of $\xi$.
The only parameters left in the model  are
the size parameters $R_\pi,R_\rho,R_\sigma$ and $R_{f_0}$  of the meson wave functions.
Assuming a common oscillator parameter for all mesons,
 meson decay analyses
yield  $R=2.5~ GeV^{-1}$ \cite{kok,ackley}. 
These values will be refered to as  set 1.
In Ref. \cite{ams}, for the two pseudoscalar meson  decay modes, 
 differences in size parameters for the $Q\bar{Q}$ states
are    taken into account,
with $R(^3P_0)=2.0 ~GeV^{-1}$ and $R(^1S_0)=2.5~ GeV^{-1}$ (set 2).
A recent full scale analysis of tensor meson decays \cite{thomas}
yields $R_{\pi}=R_{\rho}=3.69~GeV^{-1}$, 
suggesting  $R(^3P_0) \approx 3.7 GeV^{-1}$ (set 3). 
There is no   reliable constraint for $R_\sigma$;
since $\sigma$ represents a broad  $\pi\pi$ S-wave resonance, 
$R_{\sigma} \geq R_{\pi}$ is reasonable.
We resort to these  different sets of values 
to study  the sensitivity of the different observables to this input
leading   to an effective error band
of the decay model.

{\it Results} -  If we adopt
$R_\pi=R_\sigma$ and $\Gamma_{\pi^\ast}=400\,MeV$
we obtain for the 
$f_0 \rightarrow BC$ decays the following ratios:
\begin{equation}
B(\pi\pi):B(\rho\rho):B(\sigma\sigma):B(\pi^\ast\pi)
= \left\{ \begin{array}{*{5}{c}}
1&:1.62&:0.93&:0.33& \, {\rm with~ set~1}\\
1&:1.04&:0.76&:0.56&\,{\rm with~ set~2}\\
1&:0.92&:0.28&:0.19&\, {\rm with~ set~3}
\end{array}\right.
\end{equation} 
We can see that all decay rates reveal a rather strong dependence 
on the size parameters.
However, even when we vary the 
size parameters in a rather wide range, the $4\pi$ decay modes 
of the $f_0(1500)$ always follow the decay pattern
\begin{equation} \label{comp}
B(\pi^\ast\pi)\,  < B(\sigma\sigma) \, ^{<}_{\sim}B(\pi\pi) \, 
^{<}_{\sim}B(\rho\rho)
\end{equation} 
This  hierarchy of the  decay modes  remains also  unaltered when taking
into account the experimental uncertainty of $\Gamma_{\pi^\ast}$ 
by $\pm 200 MeV$. 
Similarly, a variation of the $\sigma$ size parameter 
in the range $R_\pi \le R_\sigma \le 6.0 \,GeV^{-1}$ does not 
change the order (\ref{comp}).
This  constitutes a stable prediction of the model.
Experimental data concerning the $4\pi$ decay mode of the $f_0(1500)$
and its separation into the contributions of the individual two-body channels
are still incomplete.
Recently, Crystal Barrel observed the $f_0(1500)$ in its $4\pi^0$ decay mode
in the annihilation reaction $p\bar{p} \rightarrow 5\pi^0$ at
rest \cite{CB96}.
With their results for $f_0(1500)\to \pi\pi$,
given by a single channel ($p\bar p \to 3 \pi^0$) \cite{crys} and,
more consistently, by a coupled channel analysis ($p\bar p \to \pi^0 \eta \eta
, ~\pi^0 \pi^0 \eta$ and $3\pi^0$) \cite{CB95},
the ratio r is found to be  
\begin{equation} \label{cry}
r=\frac{B(f_0(1500) \rightarrow 4\pi)}{B(f_0(1500) \rightarrow 2\pi)}=
\left\{ \begin{array}{ll}
3.4\pm0.8 & \cite{CB96} \\ 
2.1\pm0.6 & \cite{review97}
\end{array}
\right.
\end{equation}
where  the contribution of the
$\rho\rho$ intermediate state is not
included in $B(f_0(1500) \rightarrow 4\pi)$.
However, an alternative multi-channel analysis \cite{abele},
resulting in a partial decay width of $\Gamma_{\pi\pi}= 60 \pm 12~ MeV$
and a total width of $\Gamma = 132\pm 15 ~MeV$ for the $f_0(1500)$
seems to indicate a ratio of $r$ closer to 1.
For the three parameter sets and using $\Gamma_{\pi^\ast}=400~ MeV$,
$B(\pi^\ast \rightarrow \sigma\pi)=0.6$, and  $R_\sigma=R_\pi$, 
our calculation gives
\begin{eqnarray} \label{mean}
r=\frac{B(f_0 \rightarrow 4\pi)} {B(f_0 \rightarrow 2\pi)} & =&
\left\{
\begin{array}{ll}
1.13 & {\rm for~  set~1}\\
1.09 & {\rm for~  set~2}\\
0.40 & {\rm for~  set~3}
\end{array}
\right.
\end{eqnarray}
 to be compared to the experimental number of Eq. (\ref{cry}).
Again, taking into account the experimental uncertainties
in $\Gamma_{\pi^\ast}$  and in 
$B(\pi^\ast \rightarrow \sigma\pi)$ by $\pm 0.2$
the results (\ref{mean}) are changed by less than $10\%$. 
A variation of $R_\sigma$ from the value of $R_\pi $ to $6~GeV^{-1}$ leads to
ranges of predictions for r of 
$0.41\,-\,1.21$, $0.32\,-\,1.20$ and $0.26\,-\,0.45$, for the three
parameter sets respectively.
Even when one allows  strong variations in the size parameters the
theoretical predictions for r are lower than  the
experimental result by at least
a factor of two.
Although in Ref. \cite{CB96}, the final states $\sigma\sigma$ and
$\pi\pi^\ast$ contribute with similar intensities to the $4\pi^0$ 
mode, precise values for the respective
decay branching ratios cannot be given due
to interference effects.
Furthermore, preliminary results
\cite{thoma} of an analysis of $p\bar p \to 5\pi$
including charged state combinations of $\pi$,
seem to indicate that the $\sigma\sigma$ decay
is dominating over the $\rho\rho$ decay.

In summary, we have investigated  the  
glueball-quarkonia mixing
scheme of Refs. \cite{ams,wein} in the $f_0(1500)$ decay
into $\sigma\sigma, \rho\rho, \pi^\ast\pi$ and $\pi\pi$ final states.
The leading order hadronic decay mechanism, as described in Ref. \cite{ams},
proceeds both by the $Q\bar Q$ and the gluonic component of the $f_0 (1500)$;
latter process, where the glueball decays into a pair of glueballs, can
feed final states with isoscalar mesons, possessing a sizable gluonic
component.
Given the modelling of $\sigma $ as a $^3P_0$ $Q\bar Q$ state, where
gluonic components are neglected, for the leading order decay
mechanism we predict a decay pattern
of $4\pi$ modes, where the $\rho\rho$ channel dominates.
This is in conflict with preliminary experimental results \cite{thoma}, 
where the $\rho\rho$ decay mode is suppressed. 
For the $4\pi/2\pi$ decay ratio 
the result we obtain is insufficient to fully explain
the experimental result of Eq. (\ref{cry}).
Another analysis of the $f_0(1500)$ 
decays \cite{abele} seems to indicate that
the ratio of Eq. (\ref{cry}) is considerably lower,
reducing the disagreement with the theoretical results.
We stress that the decay pattern developed
here is common to a simple $n\bar{n}$ state and to the mixed $f_0(1500)$,
where its $s\bar s$ configuration does not couple to the
$4\pi$ modes.
If the experimental qualitative ordering of the
$f_0(1500)$ decay rates 
$B(4\pi)> B(2\pi)$ and $B(\sigma \sigma ) > B(\rho\rho)$ is confirmed,
the decay scheme developed here should be modified by taking into account
an additional direct coupling of the gluonic component of the $f_0 (1500)$
state to the $4 \pi$ decay channels;
for example, leading order contributions can enhance the $\sigma \sigma$
decay mode \cite{ams},
while higher order decay mechanisms can contribute to all $4 \pi$
decay channels.
Thus an accurate experimental determination of the $4\pi$
decay modes of the $f_0(1500)$ provides an additional sensitive signal for
the need to go beyond a pure $Q\bar Q$ consideration of the $f_0(1500)$,
possibly revealing the significant glueball configuration claimed to reside
in this state.  

\par
This work was supported in part by the Graduiertenkolleg "Struktur und
Wechselwirkung von Hadronen und Kernen" (DFG Mu705/3), by a grant of
the Deutsches Bundesministerium
f\"ur Bildung und Forschung (contract No. 06 T\"u 887) and by
the PROCOPE cooperation project (No. 96043).

\newpage

\begin{center}
Table: Flavor matrix elements for the decay modes $f_0\rightarrow BC$.
\end{center}

\bigskip
\bigskip

\begin{center}
\begin{tabular}{|c|c|} \hline
&\\
{\Large decay} & {\Large $\mid T_F^{(f_0 \rightarrow BC)}\mid^ 2 / 
\xi^2$} \\
&\\
\hline
&\\
{\Large $f_0 \rightarrow \pi\pi(\rho\rho)$} & {\large 3} \\ 
&\\
{\Large $f_0 \rightarrow \pi\pi^\ast$} &{\Large 6} \\ 
&\\
{\Large $f_0 \rightarrow \sigma\sigma$} &{\Large 1} \\
&\\ \hline

\end{tabular}
\end{center}

\newpage
\noindent
Figure: Decay of the $Q\bar{Q}$ component of the $f_0$ into two mesons BC.
The dot indicates the $Q\bar{Q}~^3P_0$ vertex. 
\begin{center}
\epsfxsize=15cm
\epsffile[0 50 560 730]{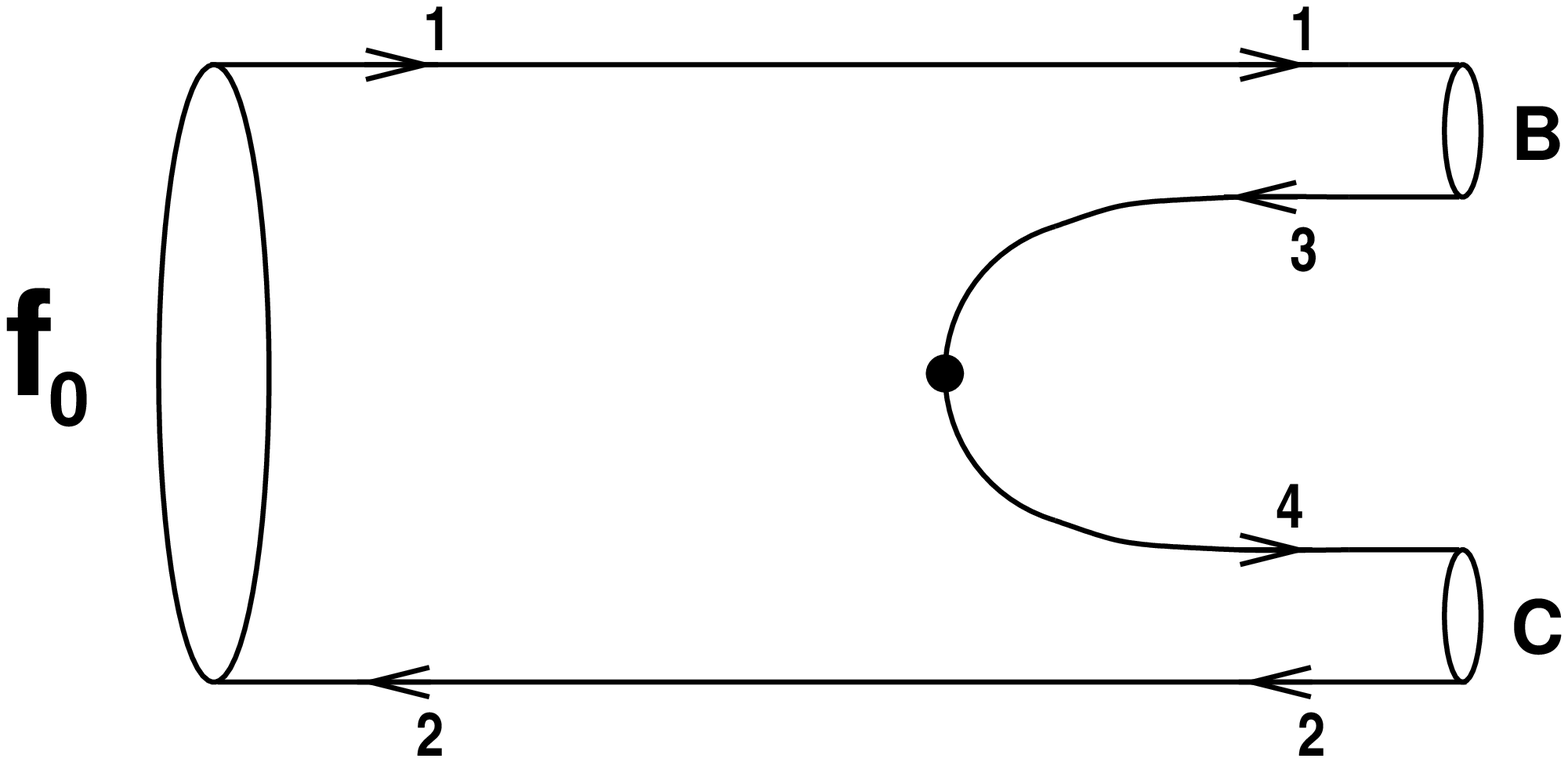}
\end{center}
\end{document}